\documentclass{PoS}

\usepackage[intlimits]{amsmath}
\usepackage{amssymb}
\usepackage{mathrsfs}
\usepackage{dsfont}
\usepackage{subfigure}

\title{Solving the sign problem of two flavor scalar electrodynamics at finite chemical potential}

\ShortTitle{Solving the sign problem of  scalar electrodynamics at finite chemical potential}

\author{Ydalia Delgado
\\Institut f\"ur Physik, Karl-Franzens Universit\"at, Graz, Austria
\\E-mail: \email{ydalia.delgado-mercado@uni-graz.at}}

\author{Christof Gattringer
\\Institut f\"ur Physik, Karl-Franzens Universit\"at, Graz, Austria
\\E-mail: \email{christof.gattringer@uni-graz.at}}

\author{Alexander Schmidt
\\Institut f\"ur Physik, Karl-Franzens Universit\"at, Graz, Austria
\\E-mail: \email{alexander.schmidt@uni-graz.at}}

\abstract{
We explore two flavor  scalar electrodynamics on the lattice, which has a 
complex phase problem at finite chemical potential. By rewriting the action 
in terms of dual variables this complex phase problem can be solved exactly. 
The dual variables are link- and plaquette occupation numbers, subject to local
constraints that have to be respected by the Monte Carlo algorithm. 
For the simulation we use a local update as well as the newly developed
``surface worm algorithm'', which is a generalization of the Prokof'ev Svistunov 
worm algorithm concept for simulating the dual representation of abelian 
Gauge-Higgs models on a lattice. We assess the performance of the two algorithms, 
present results for the phase diagram 
and discuss condensation phenomena.}

\FullConference{XXIX International Symposium on Lattice Field Theory \\
                 July 29 $-$ August 03 2013\\
                 Mainz, Germany}
 
\begin{document}

\section{Motivation}

At finite chemical potential $\mu$ the fermion determinant of QCD becomes complex
and can not be interpreted as a probability weight in a Monte Carlo simulation.
This so-called "complex phase problem" or "sign problem" has considerably 
slowed down the exploration of QCD at finite density using lattice methods.  
Although a lot of effort has been put into solving the complex phase problem of 
QCD (see, e.g., \cite{reviews} for recent reviews), the final goal of a proper ab-initio 
simulation of lattice QCD at finite density has not been achieved yet.

For some models, as well as for QCD in limiting cases, it is possible to deal with the complex phase 
problem (see, e.g., \cite{solve-sign-problem}) with different techniques. Here we use a dual 
representation, i.e., a reformulation of the system with new degrees of freedom,
which has been shown to be a very powerful method that can solve the complex 
phase problem of different models \cite{dual} without making any approximation of the partition sum.  
In the following we present another example where the dual representation can be applied successfully.  
We consider scalar QED with two flavors, i.e., a compact U(1) gauge field coupled to two complex scalar 
fields with opposite charge and a quartic self interaction \cite{prl}. We explore the full phase diagram as a 
function of the inverse gauge coupling and the mass parameter,  and present some results at finite $\mu$.

After mapping the degrees of freedom of the system to the dual variables, the weight in the 
partition sum is positive and real and usual Monte Carlo techniques can be applied.  However, 
the dual variables, links and plaquettes for this model, are subject to non-trivial constraints.
Therefore one has to choose a proper algorithm in order to sample the system efficiently.  In our case, we have
used two different Monte Carlo algorithms:  A local update algorithm \cite{z3} and an extension \cite{swa} of the
Prokof'ev Svistunov worm algorithm \cite{worm}. In addition to discussing the physics of the model, we also present
a comparison of the performance of the two algorithms

\section{Scalar electrodynamics}
 
In the conventional representation two flavor scalar electrodynamics is a model of two flavors of 
oppositely charged complex fields $\phi_x, \chi_x \in \mathds{C}$ living on the 
sites $x$ of the lattice, interacting via the gauge fields $U_{x,\sigma} \in$ U(1) sitting on the links. 
We use 4-d euclidean lattices of size $V_4 = N_s^3 \times N_t$ with periodic 
boundary conditions for all directions. The lattice spacing is set to 1, i.e., all dimensionful quantities 
are in units of the lattice spacing. 

We write the action as the sum, $S = S_U + S_\phi + S_\chi$, where $S_U$ is the gauge action 
and $S_\phi$ and $S_\chi$ are the actions for the two scalars. For the gauge action we use 
Wilson's form
\begin{equation} 
S_U \; = \; - \beta \, \sum_x \sum_{\sigma < \tau} \mbox{Re} \; U_{x,\sigma} U_{x+\widehat{\sigma}, \tau}
U_{x+\widehat{\tau},\sigma}^\star U_{x,\tau}^\star \; .
\label{gaugeaction}
\end{equation}
The sum runs over all plaquettes, $\widehat{\sigma}$ and $\widehat{\tau}$ denote the unit vectors in $\sigma$- and 
$\tau$-direction and the asterisk is used for complex conjugation.  
The action for the field $\phi$ is 
\begin{equation}
S_\phi   
\; =   \sum_x \!\Big(  M_\phi^2 \, |\phi_x|^2  + \lambda_\phi |\phi_x|^4  -
\sum_{\nu = 1}^4 \!
\big[  e^{-\mu_\phi  \delta_{\nu, 4} } \, \phi_x^\star \, U_{x,\nu} \,\phi_{x+\widehat{\nu}} 
\, + \, 
e^{\mu_\phi \delta_{\nu, 4}} \, \phi_x^\star \, 
U_{x-\widehat{\nu}, \nu}^\star \, \phi_{x-\widehat{\nu}}  \big] \!  \Big) .
\label{matteraction}  
\end{equation}
By $M_\phi^2$  we denote the combination $8 + m_\phi^2$, where $m_\phi$ is the bare mass
parameter of the field $\phi$ and $\mu_\phi$ is the chemical potential, which favors forward
hopping in time-direction (= 4-direction). We also allow for a quartic self interaction of the scalar fields and 
the corresponding coupling is denoted as $\lambda_\phi$. Note that for $\mu_\phi \neq 0$ (\ref{matteraction}) 
is complex, i.e., in the conventional form  the theory has a complex action problem.

The action for the field $\chi$ has the same form as (\ref{matteraction}) but with complex conjugate link 
variables $U_{x,\nu}$ such that $\chi$ has opposite charge.  $M_\chi^2$, $\mu_\chi$ and $\lambda_\chi$  
are used for the parameters of $\chi$. 

The partition sum $Z = \int D[U] D[\phi,\chi] e^{-S_U - S_\chi - S_\phi}$  is obtained by
integrating the Boltzmann factor over all field configurations. The measures are products over
the measures for each individual degree of freedom.

\vskip2mm
\noindent  
{\bf Dual representation:} A detailed derivation of the dual representation for the one flavor
model is given in \cite{swa} and the two flavor version we consider here simply uses two copies
of the representation of the matter fields. The dual variables for the first flavor will be denoted by 
$j_{x,\nu},  \overline{j}_{x,\nu}$, while $l_{x,\nu}$ and $\overline{l}_{x,\nu}$ are used for the second flavor.
The dual representation of the partition sum for scalar QED 
with two flavors of matter fields is given by
\begin{equation}
\hspace*{-3mm} Z = \!\!\!\!\!\! \sum_{\{p,j,\overline{j},l,\overline{l} \}} \!\!\!\!\!\!  {\cal C}_g[p,j,l]  \;  {\cal C}_s  [j] \;   {\cal C}_s  [l] \;  {\cal W}_U[p] 
\; {\cal W}_\phi \big[j,\overline{j}\,\big] \, {\cal W}_\chi \big[l,\overline{l}\,\big]  .
\label{Zfinal}
\end{equation} 
The sum runs over all configurations of the dual variables: The occupation numbers 
$p_{x,\sigma\tau} \in \mathds{Z}$ assigned to the plaquettes of the lattice and the flux variables  $j_{x,\nu},  l_{x,\nu} \in \mathds{Z}$ and
$\overline{j}_{x,\nu},  \overline{l}_{x,\nu} \in \mathds{N}_0$ living on the links. The flux variables $j$ and $l$ are subject
to the constraints ${\cal C}_s$,
\begin{equation}
 {\cal C}_s [j] \, = \, \prod_x \delta \! \left( \sum_\nu \partial_\nu j_{x,\nu}  \right)\;  , \; \;
 {\cal C}_s [l] \, = \, \prod_x \delta \! \left( \sum_\nu \partial_\nu l_{x,\nu}  \right) , \;
\label{loopconstU1}
\end{equation}
which enforce the conservation of $j$-flux and of $l$-flux at each site of the lattice
(here $\delta(n)$ denotes the Kronecker delta $\delta_{n,0}$ and $\partial_\nu f_x \equiv 
f_x - f_{x-\widehat{\nu}}$).
Another constraint,
\begin{equation}
 {\cal C}_g [p,j,l]  \! =\!  \prod_{x,\nu} \! \delta  \Bigg( \!\sum_{\nu < \alpha}\! \partial_\nu p_{x,\nu\alpha}  
- \!\sum_{\alpha<\nu}\! \partial_\nu p_{x,\alpha\nu} + j_{x,\nu} - l_{x,\nu} \! \Bigg)\! ,
\label{plaqconstU1}  
\end{equation}
connects the plaquette occupation numbers $p$ with the $j$- and $l$-variables. 
At every link it enforces the combined flux of the plaquette occupation 
numbers  plus the difference of $j$- and $l$-flux residing on that link to vanish.  The
fact that $j$- and $l$-flux enter with opposite sign is due to the opposite charge of the two 
flavors.

The constraints (\ref{loopconstU1}) and (\ref{plaqconstU1}) restrict the admissible
flux and plaquette occupation numbers giving rise to an interesting geometrical
interpretation: The $j$- and $l$-fluxes form closed oriented loops made of links. The integers
$j_{x,\nu}$ and $l_{x,\nu}$ determine how often a link is run through by loop segments, with negative
numbers indicating net flux in the negative direction. The flux conservation 
(\ref{loopconstU1}) ensures that only closed loops appear. Similarly, the constraint 
(\ref{plaqconstU1}) for the plaquette occupation numbers can be seen as a continuity
condition for surfaces made of plaquettes. The surfaces are either closed
surfaces without boundaries or open surfaces bounded by  $j$- or $l$-flux.

The configurations of plaquette occupation numbers and fluxes in (\ref{Zfinal}) come with 
weight factors 
\begin{eqnarray}
{\cal W}_U[p] & = & \!\! \! \prod_{x,\sigma < \tau} \! \! \!
 I_{p_{x,\sigma\tau}}(\beta) \, ,
\\   
{\cal W}_\phi \big[j,\overline{j}\big] & = & 
\prod_{x,\nu}\! \frac{1}{(|j_{x,\nu}|\! +\! \overline{j}_{x,\nu})! \, 
\overline{j}_{x,\nu}!} 
\prod_x e^{-\mu j_{x,4}}  P_\phi \left( f_x \right) ,
\nonumber
\end{eqnarray}
with 
\begin{equation}
f_x \; = \; \sum_\nu\!\big[ |j_{x,\nu}|\!+\!  |j_{x-\widehat{\nu},\nu}|  \!+\!
2\overline{j}_{x,\nu}\! +\! 2\overline{j}_{x-\widehat{\nu},\nu} \big] \; ,
\end{equation}
which is an even number. The $I_p(\beta)$
in the weights  ${\cal W}_U$ are the modified Bessel functions and the $P_\phi (2n)$ in 
${\cal W}_\phi$  are the integrals
\begin{equation}
P_\phi (2n)  \; = \;  \int_0^\infty dr \, r^{2n+1}
\,  e^{-M_\phi^2\, r^2 - \lambda_\phi r^4} = \sqrt{\frac{\pi}{16 \lambda}}  \, \left(\frac{-\partial}{\partial M^2}\right)^{\!n} \;  
e^{\, M^4 / 4 \lambda} \left[1- erf(M^2/2\sqrt{\lambda})\right] \; .
\end{equation}
These integrals are related to derivatives of the error function and we evaluate them numerically and
pre-store them for the Monte Carlo simulation. The weight factors $ {\cal
W}_\chi$ are the same as the $ {\cal W}_\phi$, only  the parameters $M_\phi^2$,
$\lambda_\phi$, $\mu_\phi$ are replaced by  $M_\chi^2$, $\lambda_\chi$, $\mu_\chi$. All
weight factors are real and positive and the partition sum (\ref{Zfinal}) thus  is
accessible  to Monte Carlo techniques,  using the plaquette occupation numbers and the
flux variables as the new degrees of freedom.

\section{Monte Carlo simulation}
 
Because the dual variables are subject to non-trivial constraints, they cannot be modified randomly during 
the update. Here we use two strategies: A local update, referred to as LMA (local Metropolis algorithm),
which consists of three types of steps: Steps where we change plaquettes bounded by matter flux,  steps where 
we change the plaquettes on 3-cubes, and steps where we propose double lines of matter flux around the temporal 
direction. These changes are built such that the constraints remain intact for each individual step and the 
tests of the LMA are reported in \cite{prl,z3,swa}.  

Another possibility is to use an extension of the worm
algorithm \cite{worm}, the so called surface worm algorithm \cite{swa}, which we refer to as SWA. Here initially 
the constraints are violated at a single link and the SWA subsequently propagates this defect on the lattice 
until the defect is healed in a final step. For both the LMA and the SWA the unconstrained $\overline{l}$ and 
$\overline{j}$ variables are updated with conventional Metropolis steps. 
Here we present results for both algorithms and
assess their performance.

\subsection{Local Metropolis algorithm LMA}
\begin{figure}[h]
\begin{center}
\includegraphics[width=\textwidth,clip]{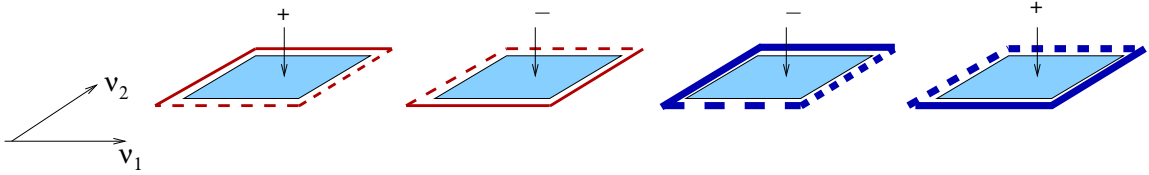}
\end{center}
\vspace{-4mm}
\caption{Plaquette update: A plaquette occupation number is changed by $+1$ or
$-1$ and the links $j$ (thin red links) or $l$ (fat blue links) of the plaquette are changed simultaneously. The
full line indicates an increase by +1 and a dashed line a decrease by $-1$. 
The directions $1 \le \nu_1 < \nu_2 \le 4$
indicate the plane of the plaquette.} \label{plaquette}
\vspace{-2mm}
\end{figure}

\begin{figure}[h]
\begin{center}
\includegraphics[width=0.7\textwidth,clip]{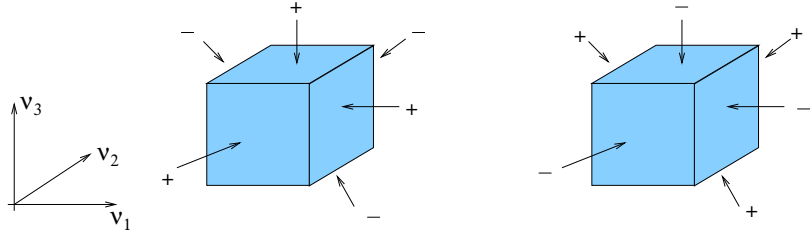}
\end{center}
\vspace{-4mm}
\caption{Cube update: Here we show the two possible changes in the plaquette occupation numbers on a 3-cube. 
The edges of the 3-cube are parallel to the directions $1 \leq \nu_1 < \nu_2 < \nu_3 \leq 4$.} \label{cube}
\vspace*{-2mm}
\end{figure}

Let us begin by describing the LMA. It consists of the following update steps:
\begin{itemize}
\vspace*{-1mm}
\item A sweep for each unconstrained variable $\overline{l}$ and $\overline{j}$ 
raising or lowering their occupation number by one unit.
\vspace*{-1mm}
\item ``Plaquette update'': 
It consists of increasing or decreasing a plaquette occupation number
$p_{x,\nu\rho}$ and
the link fluxes (either $j_{x,\sigma}$ or $l_{x,\sigma}$) at the edges of $p_{x,\nu\rho}$ by $\pm 1$ as 
illustrated in Fig.~\ref{plaquette}. The change of $p_{x, \nu \rho}$ 
by $\pm 1$ is indicated by the signs $+$ or $-$, while the flux variables $j$ ($l$) are denoted by the thin red line
(fat blue lines for the second flavor) and we use a dashed line to indicate a decrease by $-1$ and a full line 
for an increase by $+1$.
\vspace*{-1mm}
\item ``Winding loop update'': 
It consists of increasing or decreasing the occupation number of both link variables $l$ and $j$ by 
one unit along a winding loop in any of the 4 directions.  This update is very important because the winding loops
in time direction are the only objects that couple to the chemical potential.
\vspace*{-1mm}
\item ``Cube update'':  The plaquettes of 3-cubes
of our 4-d lattice are changed according to one of the two patterns illustrated in 
Fig.~\ref{cube}. 
Although the plaquette and winding loop update are enough to satisfy ergodicity, 
the cube update helps for decorrelation in the region of 
parameters where the system is dominated by closed surfaces, i.e., where the link
acceptance rate is small.
\end{itemize}
\vspace*{-1mm}
A full sweep consists of updating all links, plaquettes, 3-cubes and winding loops on the lattice,
offering one of the changes mentioned above and accepting them with the Metropolis 
probability computed from the local weight factors.

\subsection{Surface worm algorithm SWA}
 
Instead of the LMA we can use a generalization of the the worm algorithm, the SWA.
Here we only shortly describe the SWA and refer to  \cite{swa} for a detailed description.

\begin{figure}[h]
\begin{center}
\includegraphics[width=\textwidth,clip]{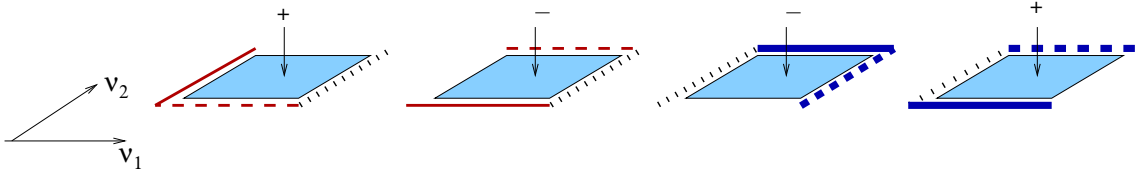}
\end{center}
\vspace{-4mm}
\caption{Examples of segments for the links $j$ (lhs.) and $l$ (rhs.) 
in the $\nu_1$-$\nu_2$-plane ($\nu_1 < \nu_2$).
The plaquette occupation numbers are changed as indicated by the signs. 
The full (dashed) links are changed by $+1$ ($-1$). The empty link shows
where the segment is attached to the worm and the dotted link is the new position of the link
$L_V$ where the constraints are violated.} \label{segments}
\vspace{-2mm}
\end{figure}

\begin{figure}[h]
\begin{center}
\includegraphics[width=\textwidth,clip]{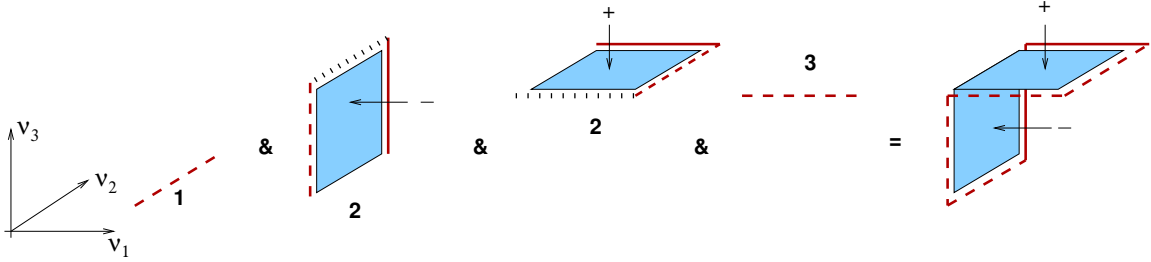}
\end{center}
\vspace{-4mm}
\caption{A simple example for an update with the surface worm algorithm.  
See the text for a detailed explanation of the steps involved.} \label{worm}
\vspace{-2mm}
\end{figure}

The SWA is constructed by breaking up the smallest update elements of the LMA, i.e., the plaquette updates, 
into smaller building blocks called ``segments'' (examples are shown in Fig.~\ref{segments}), used to build 
larger surfaces on which the flux and plaquette variables are changed. In the SWA the constraints are temporarily 
violated at a link $L_V$, the head of the worm, and the two sites at its endpoints. The SWA then transports this defect on the
lattice until it closes with a final step that heals the constraint.
The admissible configurations are generated using 3 elements:

\begin{enumerate}

\item The worm starts by changing either the $l$ or the $j$  flux by $\pm 1$ at 
a randomly chosen link (step 1 in Fig.~\ref{worm} where a worm for $j$ fluxes starts).

\item The first link becomes the head of the worm $L_V$.
The defect at $L_V$ is then propagated through the lattice by 
attaching segments of the same kind of flux ($j$ or $l$) as the first segment, 
which are chosen in such a way that the constraints are always 
obeyed at the link where the next segment is attached (step 2 in Fig.~\ref{worm}).

\item Attaching segments the defect is propagated through the lattice until the worm decides to
end with the insertion of another unit of link flux at $L_V$ (step 3 in Fig.~\ref{worm}) to heal the violated constraint.
 
\end{enumerate}
A full sweep consists of $V_4$ worms with $l$ fluxes and $V_4$ worms with $j$ fluxes, 
plus a sweep of the unconstrained 
variables $\overline{l}$ and $\overline{j}$,
and a sweep of winding loops (as explained for the LMA).

\section{Results}
\vspace{-1mm}
\noindent
In this section we discuss the results from the numerical analysis.  We first show 
an assessment of both algorithms and compare their performance. Subsequently 
we discuss the physics of scalar QED at finite density and present the phase diagram.  
In both cases we use thermodynamical observables and their fluctuations. In particular 
we use the following observables which can be evaluated as simple derivatives of 
$\ln Z$ in both the conventional and the dual representations:  

\vskip5mm
\noindent
The first and second derivatives with respect to the inverse gauge coupling $\beta$, i.e., 
the plaquette expectation value and its susceptibility,

\begin{equation}
\langle U \rangle = \frac{1}{6 N_s^3 N_t}\frac{\partial}{\partial \beta} \ln\ Z\quad , \quad
\chi_{U} = \frac{1}{6 N_s^3 N_t}\frac{\partial^2}{\partial \beta^2} \ln\ Z\ .
\end{equation}

\noindent We also consider the particle number density $n$ 
and its susceptibility which are the first and second derivatives 
with respect to the chemical potential,

\begin{equation}
\langle n \rangle  = \frac{1}{N_s^3 N_t}\frac{\partial}{\partial \mu} \ln\ Z\quad , \quad
\chi_{n} = \frac{1}{N_s^3 N_t}\frac{\partial^2}{\partial \mu^2} \ln\ Z\ .
\end{equation}

\noindent Finally, we analyze the derivatives with respect to $M^2$,

\begin{equation}
\langle |\phi|^2 \rangle = \frac{1}{N_s^3 N_t}\frac{\partial}{\partial M^2} \ln\ Z\quad , \quad
\chi_{|\phi|^2} = \frac{1}{N_s^3 N_t}\frac{\partial^2}{\partial (M^2)^2} \ln\ Z\ .
\end{equation}

\subsection{Assessment of the LMA and SWA algorithms}
\noindent
For the comparison of our two algorithms we considered the U(1) gauge-Higgs model coupled
with one (see \cite{swa}) and two scalar fields (as described here).  
First we checked the correctness of the SWA comparing the results for different 
lattices sizes and parameters.  Examples for the one flavor model were presented  
in \cite{swa}. 

\begin{figure}[h]
\begin{center}
\hbox{\includegraphics[width=\textwidth,clip]{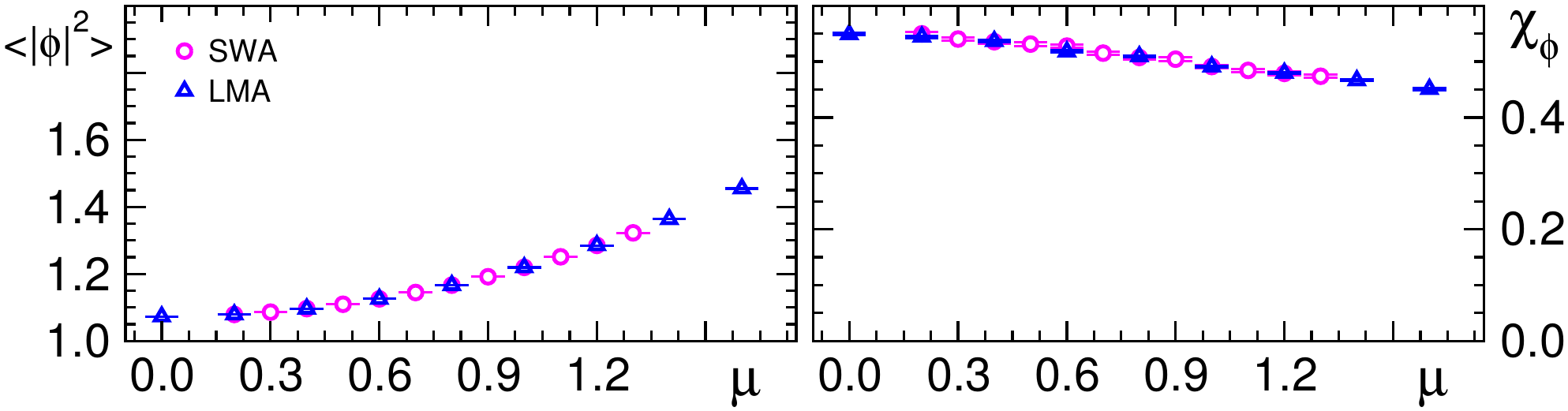}}
\vskip5mm
\hbox{\hspace{4mm}\includegraphics[width=0.97\textwidth,clip]{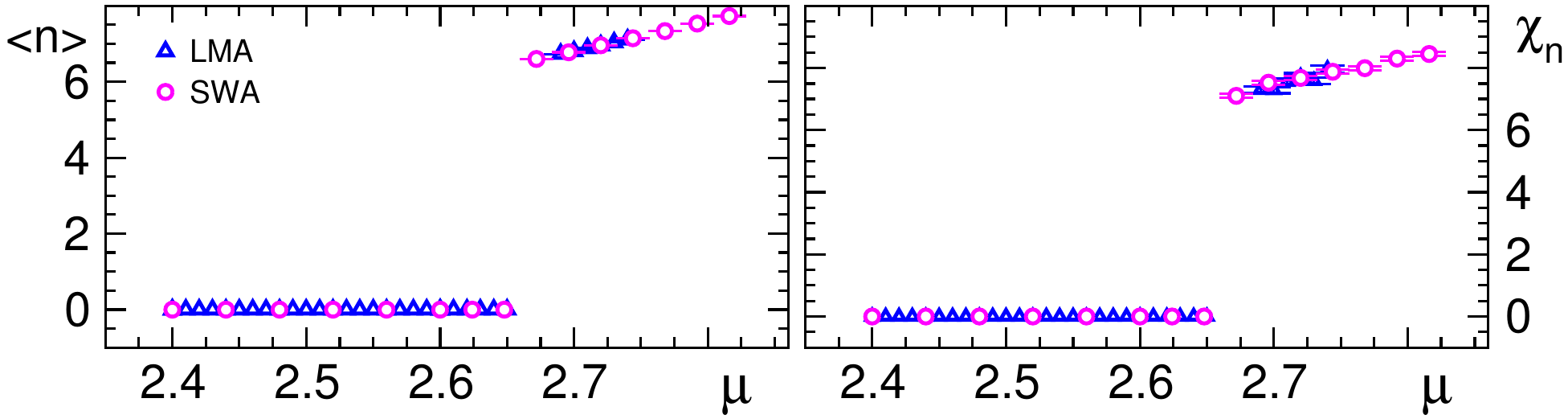}}
\end{center}
\vspace{-6mm}
\caption{Observables for the two flavor model as a function of $\mu$ for different 
parameters on a $12^3 \times 60$ lattice.
We compare results from the SWA (circles) and the LMA (triangles).} \label{obs}
\vspace*{-2mm}
\end{figure}

In Fig.~\ref{obs} we now show some examples for the two flavor case. The top figures 
of Fig.~\ref{obs} show 
$\langle |\phi|^2 \rangle$ (lhs.) and the corresponding susceptibility (rhs.) as a function of 
$\mu_\phi = \mu_\chi = \mu$ at $\beta = 0.85$ and 
$M_\phi^2 = M_\chi^2 = M^2  = 5.325$ on a lattice of size $12^3 \times 60$.  This point is located
in the Higgs phase and does not show any phase transition as a function of $\mu$.  The bottom
plots show the particle number $\langle n \rangle$ (lhs.) and its susceptibility (rhs.) as a function of $\mu$
for $\beta = 0.75$ and $M^2 = 5.73$ on a lattice of volume $12^3 \times 60$.  Here we observe
a  pronounced first order transition from the confining phase into the Higgs phase. 
It is obvious that in all four plots the agreement between the results from the LMA and from the 
SWA is excellent.

\begin{figure}[t]
\begin{center}
\includegraphics[width=\textwidth,clip]{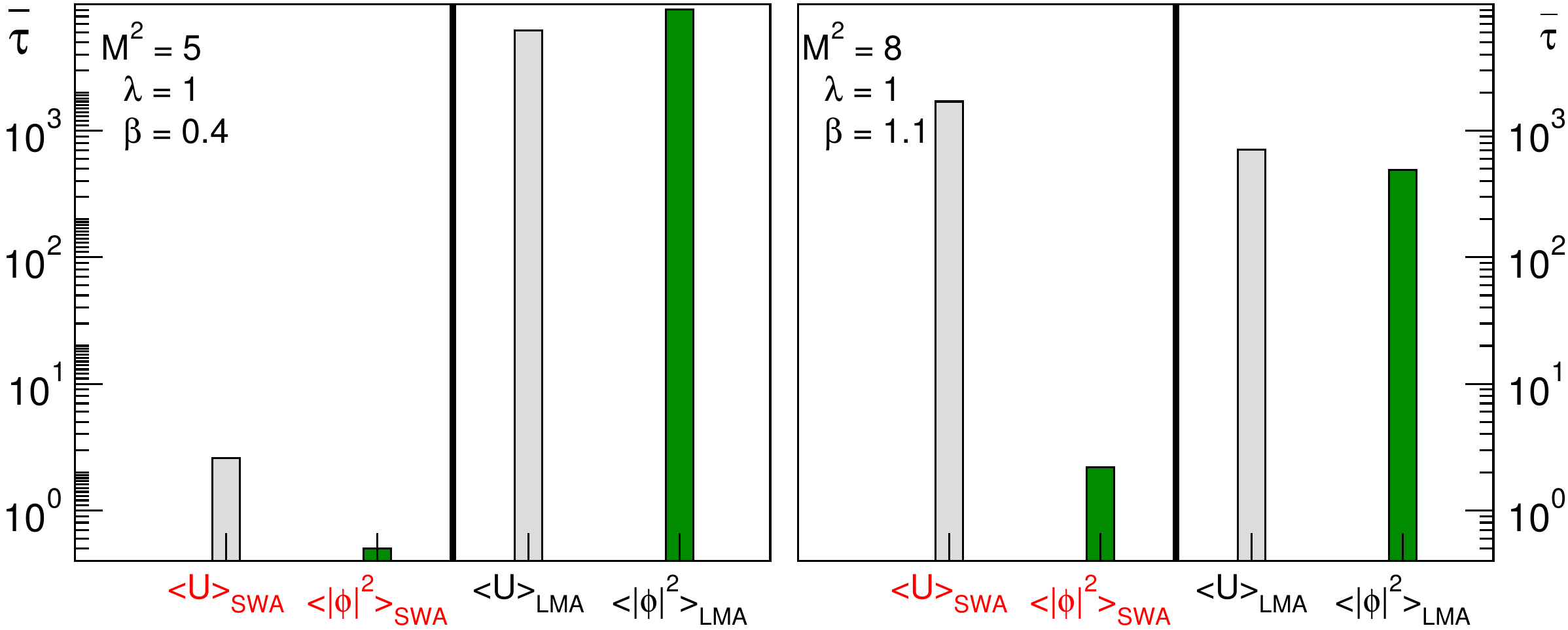}
\end{center}
\vspace{-4mm}
\caption{Normalized autocorrelation times $\overline{\tau}$ for the observables $\langle U \rangle$ and 
$\langle |\phi|^2 | \rangle$ for two different sets
of parameters for the one flavor model.  Left: Parameter values close to a first order phase transition. 
Right: A parameter set characterized by a low acceptance for matter flux.  Both simulations  
were done on $16^4$ lattices, with data taken from \cite{swa}.} \label{auto}
\vspace*{-2mm}
\end{figure}

\noindent
In order to obtain a measure of the computational effort, in \cite{swa} we compared the normalized 
autocorrelation time $\overline{\tau}$ of the SWA and LMA for 
the one flavor model for different volumes and parameters.  We concluded that,
the SWA outperforms the local update near a phase transition and if
the acceptance rate of the constrained link variables is not very low (e.g., lhs.\ of Fig.~\ref{auto}).  
On the other hand, for parameter values where the constrained links have a very low acceptance rate 
the worm algorithm has difficulties to efficiently sample the 
system because it changes the link occupation number in every move, while the LMA has a sweep with only
closed surfaces. The plot on the rhs. of Fig.~\ref{auto} shows how $\overline{\tau}$ for
$\langle U \rangle$ is larger for the SWA than for the LMA.  We remark however, that this performance issue
can be overcome easily by augmenting the SWA with sweeps of cube updates as used in the LMA.

\subsection{Physics results}
So far one of the main physics results of our studies of 2-flavor scalar QED  
(already published in \cite{prl}) is the full phase diagram of the considered 
model in the $\beta$-$M^2$ plane (using $M_\phi^2 = M_\chi^2 = M^2$)
at $\mu=0$ and the analysis of phase 
transitions driven by the chemical potential $\mu_\phi = \mu_\chi = \mu$ 
when starting from the different 
phases of the model. For the sake of completeness we here again show the 
$\mu = 0$ phase diagram, and then present new results for the observables 
in the $\beta$-$M^2$ plane at several values of $\mu > 0$, which illustrate the 
shift of the phase-boundaries at $\mu >  0$, i.e., the positions of the critical surfaces.
In addition we show that some of the transitions at finite $\mu$ can be seen as
condensation phenomena of the dual occupation numbers.

\subsubsection*{Phase diagram at $\mu=0$}

\begin{figure}[h]
\centering
\hspace*{-3mm}
\includegraphics[width=85mm,clip]{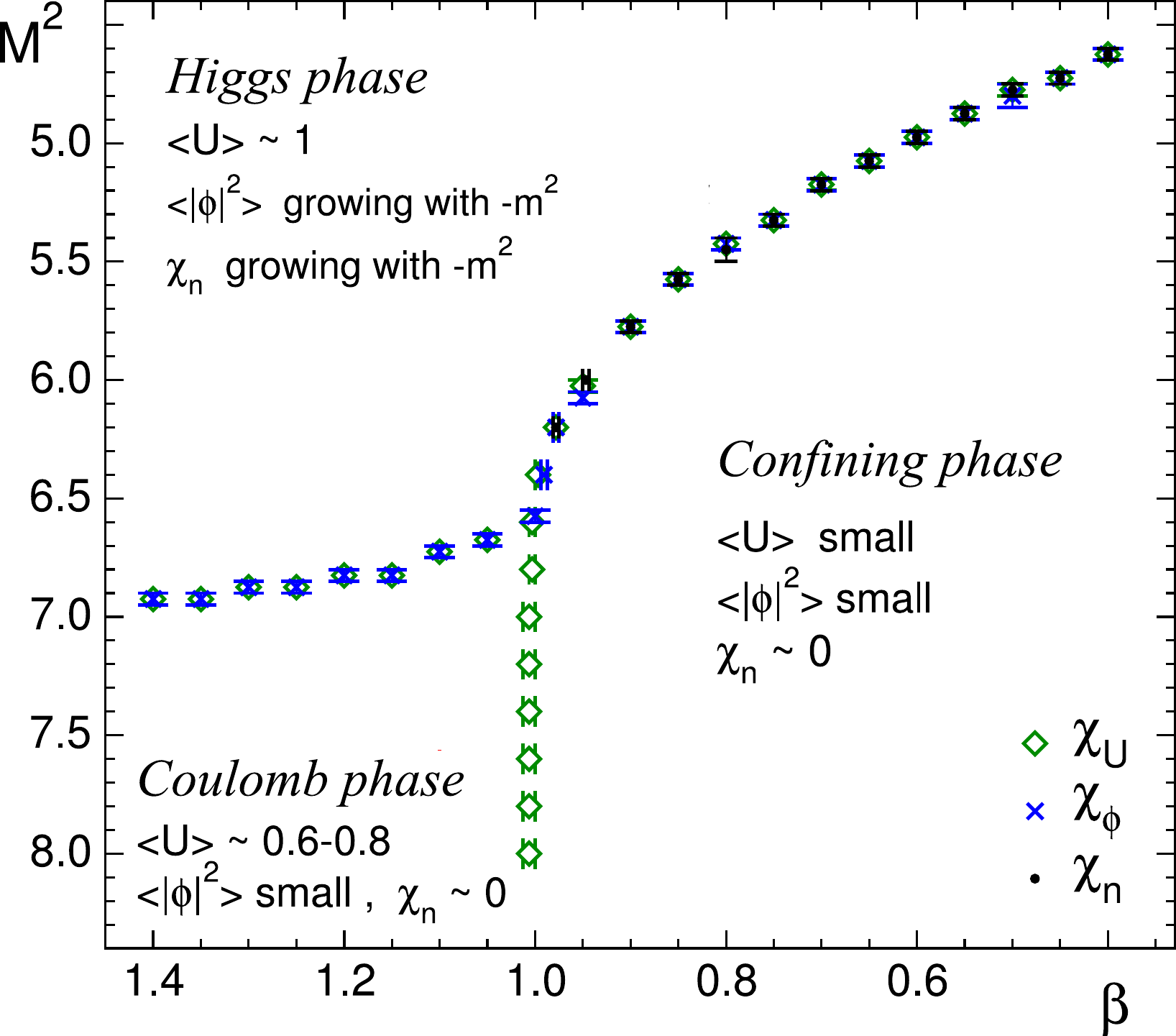}
\caption{Phase diagram in the $\beta$-$M^2$ plane at $\mu = 0$. We show
the phase boundaries determined from the maxima of the susceptibilities $\chi_U$ and $\chi_{\phi}$ and the
inflection points of $\chi_n$.}
\label{phasediagram}
\end{figure}

The results for the phase diagram at $\mu = 0$ are summarized in Fig.~\ref{phasediagram}. The various phase 
boundaries were determined from the observables $\langle U \rangle$ and $\langle |\phi|^2 \rangle$ and the 
corresponding susceptibilities. We found that the phase boundary separating  Higgs- and
confining phase is of strong first order, the line separating confining- and Coulomb phase is  of weak
first order, and the boundary between Coulomb- and Higgs phase is a continuous transition. 
Our results for the $\mu = 0$ phase diagram are in qualitative
agreement with the results for related
models \cite{Lang} studied in the conventional formulation.

\begin{figure}[p]
\centering
\hspace*{-3mm}
\includegraphics[width=\linewidth,clip]{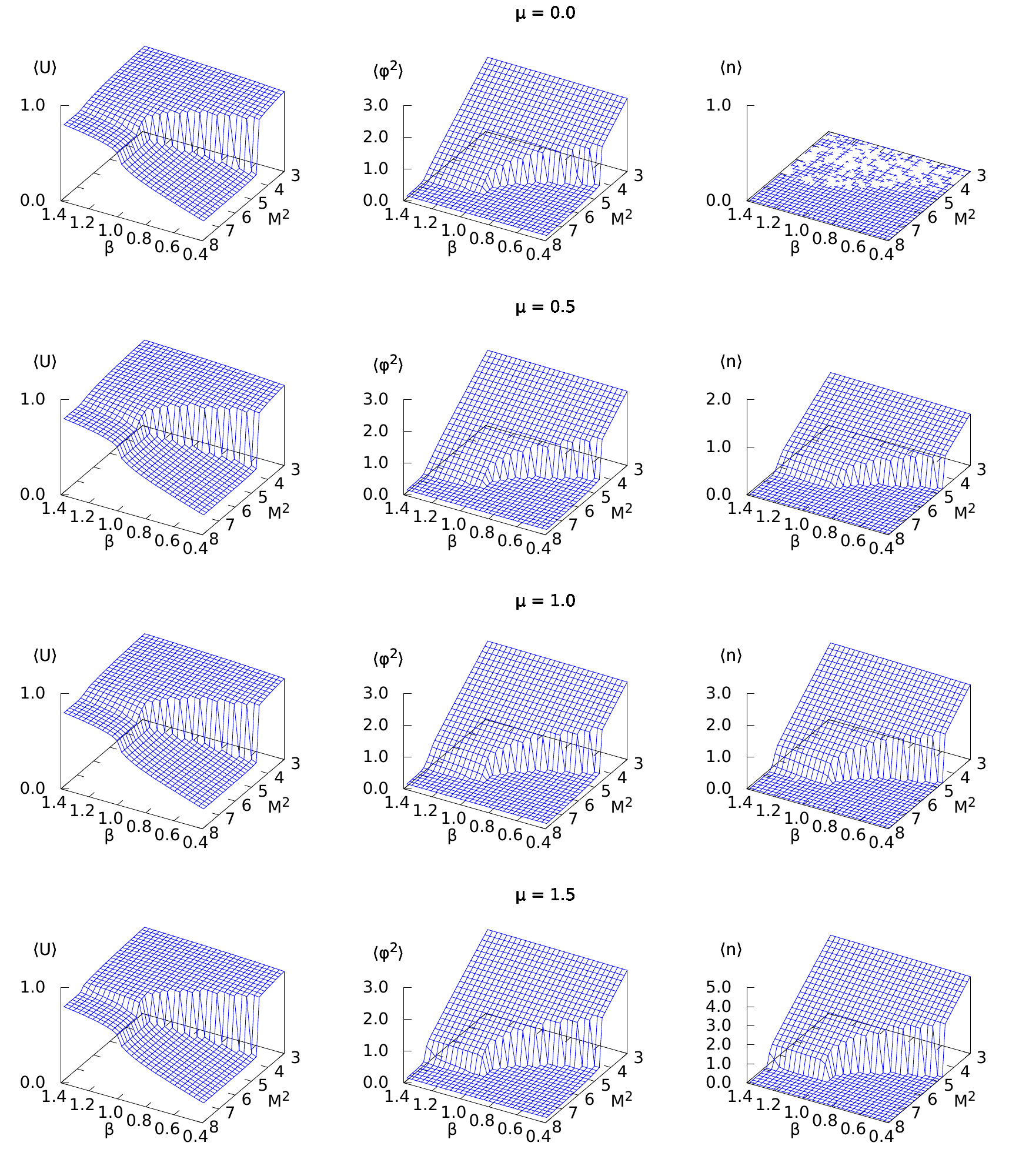}
\caption{The observables $\langle U \rangle$, $\langle |\phi|^2 \rangle$, and 
$\langle n \rangle$ as a function of $\beta$ and $M^2$ for different chemical 
potentials $\mu = 0.0,\,0.5,\,1.0$ and $1.5$. It can be seen how the phase 
boundaries shift with increasing chemical potential.}
\label{muphases}
\end{figure}

\begin{figure}[t]
\centering
\hspace*{-3mm}
\includegraphics[width=\linewidth,clip]{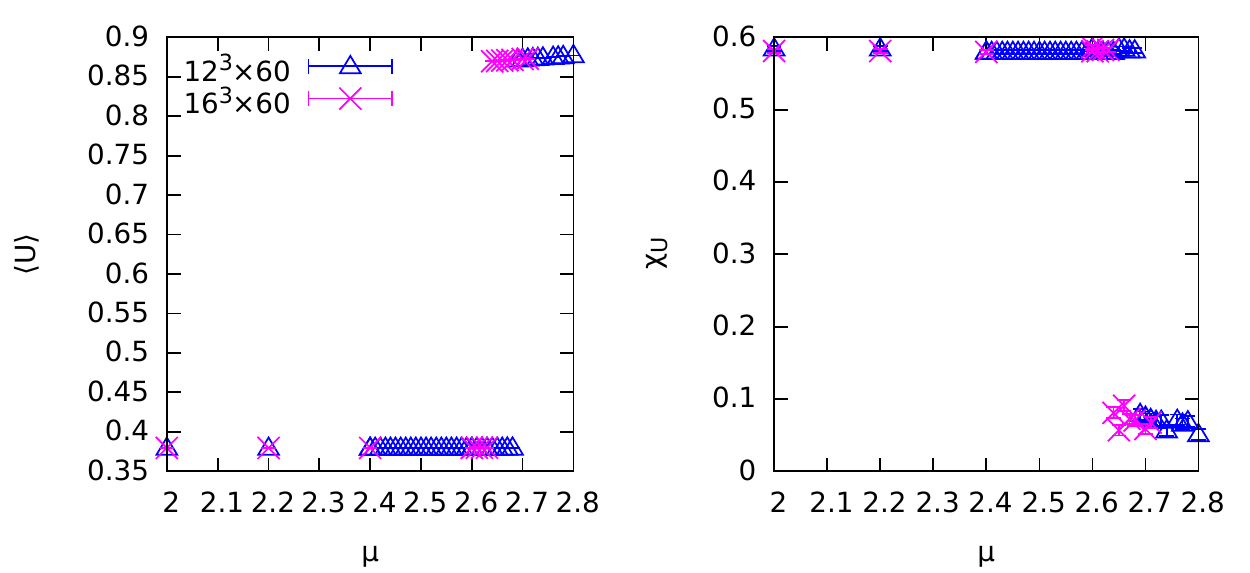}
\caption{We here show the plaquette expectation value $\langle U \rangle$ and the corresponding suscpetibility $\chi_U$ as function of the chemical potential, for two different volumes $12^3\times60$ and $16^3\times60$.}
\label{occutrans_plaq}
\end{figure}
\begin{figure}[b]
\centering
\hspace*{-3mm}
\includegraphics[width=\linewidth,clip]{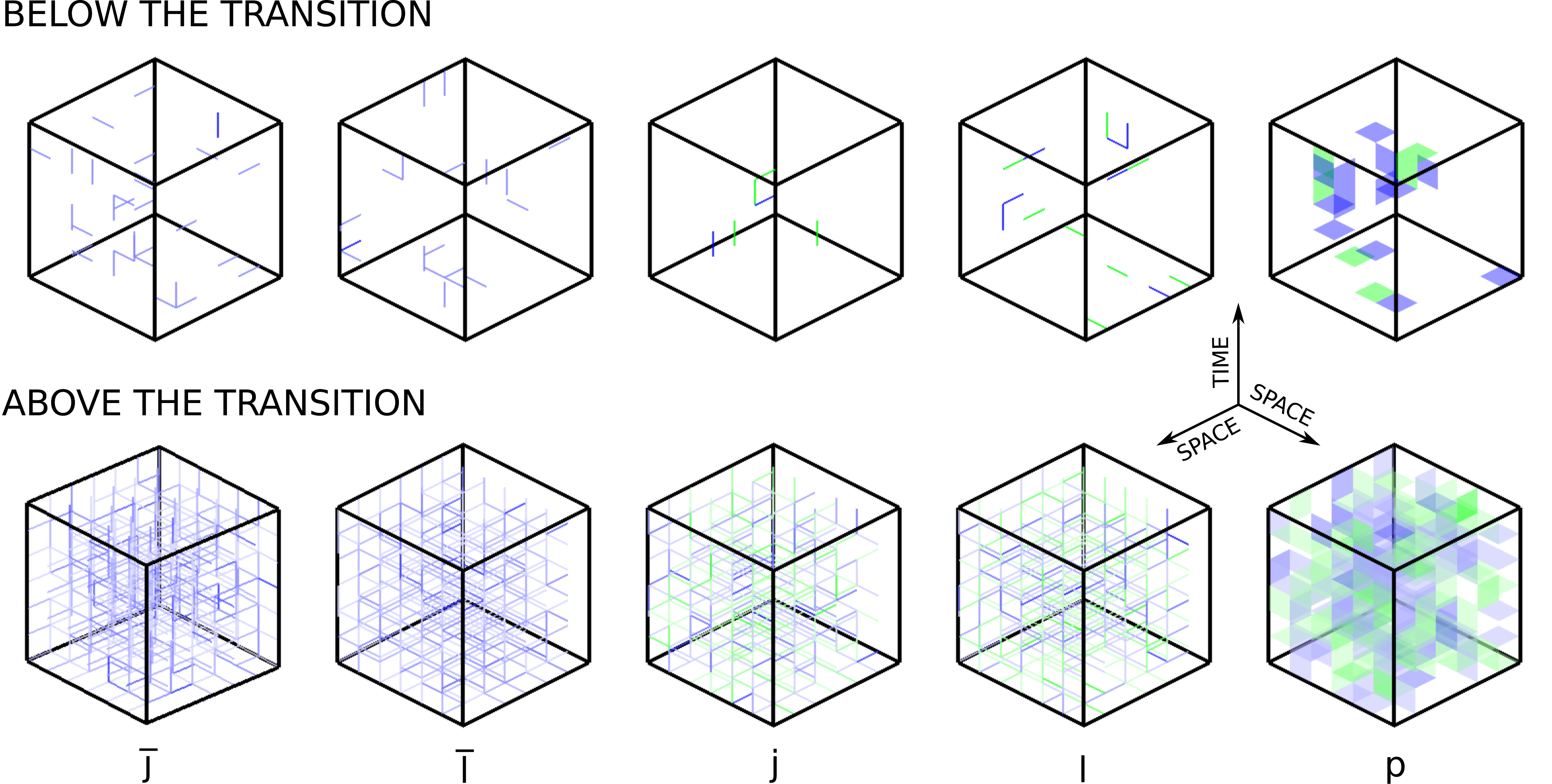}
\caption{Link occupation numbers $\bar{j}$, $\bar{l}$, $j$, $l$ and plaquette occupation numbers $p$ for values of $\mu$
just below (top) and above (bottom) the critical value $\mu_c$ for the transition from the confining- to the Higgs-phase.}
\label{occutrans}
\end{figure}

\subsubsection*{Phase boundaries at $\mu > 0$}

As a first step in the determination of the phase boundaries as functions of all three parameters $\beta, \, M^2$ and $\mu$, 
in Fig.~\ref{muphases} we plot the observables $\langle U \rangle$, $\langle |\phi|^2 \rangle$ and $\langle n \rangle$ as functions 
of $\beta$ and $M^2$ for four different values of the chemical potential $\mu=0.0,\, 0.5,\, 1.0$ and $1.5$.

The phase-transition from the confining phase to the Coulomb phase shown in Fig.~\ref{phasediagram} 
is characterized by a rapid increase of  $\langle U \rangle$ across the transition but does not give rise to 
significant changes in the other observables (compare the top row of plots in Fig.~\ref{muphases}). 
This behavior persists also at finite $\mu$ and  the 
confinement-Coulomb transition can only be seen in the $\langle U \rangle$-plots.

The transition between the Higgs- and the confinig phase is characterized by a strong first order discontinuity in all observables 
(except for $\langle n \rangle = 0$ at $\mu = 0$), a feature that persists for all our values of $\mu$. Also the transition between the Higgs- and the 
Coulomb phase is seen in all observables. It is obvious from the plots, that with increasing $\mu$ all three transitions become more pronounced in 
all variables they are seen in, and the Higgs-Coulomb transition might even change from crossover to first order. Still, the shown results 
have to be considered preliminary and more detailed studies will be necessary to draw final conclusions. 

\subsubsection*{Dual occupation numbers}
 
The dual reformulation of lattice field theories makes it possible to look at the same physics from a different perspective
by studying the dynamics of the dual degrees of freedom instead of the conventional ones. 
This being a feature we find especially interesting about the dual formulation, we here present an example where a transition 
manifests itself as the condensation of dual variables.
 
Let us first look at the transition using the standard observables. In Fig.~\ref{occutrans_plaq} we 
plot the plaquette expectation value $\langle U \rangle$ and the corresponding susceptibility $\chi_U$ 
as function of the chemical potential, for two different volumes $12^3\times60$ and $16^3\times60$. 
We see that for the larger volume the transition is shifted slightly towards lower chemical potential, 
but the volume dependence seems to be reasonably small. The parameters $\beta$ and $M^2$ are 
fixed to $\beta=0.75$ and $M^2=5.73$. Increasing the chemical potential takes us from the confining- 
to the Higgs phase where we cross the phase boundary 
at some critical value of $\mu$, which is $\mu_c\simeq2.65$ 
for the larger of the two lattices. Below the critical value of the chemical potential both 
$\langle U \rangle$ and $\chi_U$ are independent of $\mu$, which is characteristic for a Silver Blaze type of transition \cite{cohen}.
At $\mu_c$ a strong first order transition signals the entry into the Higgs phase.
 
In Fig.~\ref{occutrans} we have a look at the same transition, by now showing typical configurations of the dual variables 
just below (top) and above (bottom) the critical chemical potential $\mu_c$.
In particular we show snapshots of the occupation numbers of all dual link variables $\bar{j}$, $\bar{l}$, $j$, 
$l$ and dual plaquette variables $p$.  Here blue links/plaquettes depict positive occupation numbers, 
green links/plaquettes depict negative occupation numbers and links/plaquettes with $0$-occupation 
are not shown. It can be seen that below $\mu_c$ links and plaquettes are hardly occupied, 
while above $\mu_c$ their occupation is abundant. In that sense the Silver Blaze transition of Fig.~\ref{occutrans_plaq} 
can be understood as a condensation phenomenon of the dual variables, which is a new perspective on the underlying 
physics we gained from the dual reformulation of the problem.

\section*{Acknowledgments} 
\noindent
We thank Hans Gerd Evertz 
for numerous discussions that helped to shape this project and for 
providing us with the software to compute the autocorrelation times. 
We also acknowledge interesting discussions with Thomas Kloiber 
on aspects of the dual formulation for charged scalar fields. 
This work was supported by the Austrian Science Fund, 
FWF, DK {\it Hadrons in Vacuum, Nuclei, and Stars} (FWF DK W1203-N16). Y.~Delgado is supported by
the Research Executive Agency (REA) of the European Union 
under Grant Agreement number PITN-GA-2009-238353 (ITN STRONGnet) and by {\it Hadron Physics 2}. 
Furthermore this work is partly supported by DFG TR55, ``{\sl Hadron Properties from Lattice QCD}'' 
and by the Austrian Science Fund FWF Grant.\ Nr.\ I 1452-N27.


\begin{thebibliography}{123456}
\bibitem{reviews}
  P.~Petreczky,
  PoS ConfinementX {\bf } (2012) 028
  [arXiv:1301.6188 [hep-lat]].
%
  G.~Aarts,
  PoS LATTICE {\bf 2012} (2012) 017
  [arXiv:1302.3028 [hep-lat]].
  
\bibitem{solve-sign-problem}
  D.~Sexty,
  arXiv:1307.7748 [hep-lat].
%
  S.~Chandrasekharan,
  Eur.\ Phys.\ J.\ A {\bf 49} (2013) 90
  [arXiv:1304.4900 [hep-lat]].
%
  G.~Aarts, P.~Giudice, E.~Seiler and E.~Seiler,
  Annals Phys.\  {\bf 337} (2013) 238
  [arXiv:1306.3075 [hep-lat]].
%
  G.~Aarts, L.~Bongiovanni, E.~Seiler, D.~Sexty and I.~-O.~Stamatescu,
  Eur.\ Phys.\ J.\ A {\bf 49} (2013) 89
  [arXiv:1303.6425 [hep-lat]].
%
  M.~Cristoforetti, F.~Di Renzo, A.~Mukherjee and L.~Scorzato,
  Phys.\ Rev.\ D {\bf 88} (2013) 051501
  [arXiv:1303.7204 [hep-lat]].
%
  J.~Bloch,
  J.\ Phys.\ Conf.\ Ser.\  {\bf 432} (2013) 012023.
%
  M.~Fromm, J.~Langelage, S.~Lottini, O.~Philipsen,
  JHEP {\bf 1201} (2012) 042.
%
  M.~Fromm, J.~Langelage, S.~Lottini, M.~Neuman, O.~Philipsen,
  Phys.\ Rev.\ Lett. 110 (2013) 122001.
%
  K.~Langfeld, B.~Lucini and A.~Rago,
  Phys.\ Rev.\ Lett.\  {\bf 109} (2012) 111601
  [arXiv:1204.3243 [hep-lat]].
  
  
\bibitem{dual}
  A.~Patel, Nucl.~Phys. B {\bf 243} (1984) 411;
  Phys.\ Lett.\  B {\bf 139} (1984) 394.
  %
  T.~DeGrand and C.~DeTar, 
  Nucl.\ Phys.\  B {\bf 225} (1983) 590. 
  %
  J.~Condella and C.~DeTar,
  Phys.\ Rev.\  D {\bf 61} (2000) 074023,
  [arXiv:hep-lat/9910028].
%
 C.~Gattringer and T.~Kloiber,
  Nucl.\ Phys.\ B {\bf 869} (2013) 56
  [arXiv:1206.2954 [hep-lat]].
  C.~Gattringer and T.~Kloiber,
  Phys.\ Lett.\ B {\bf 720} (2013) 210
  [arXiv:1212.3770 [hep-lat]].
%
  T.~Sterling, J.~Greensite,
  Nucl.\ Phys.\ B {\bf 220} (1983) 327.
%
  M.~Panero,
  JHEP {\bf 0505} (2005) 066.
%
  V.~Azcoiti, E.~Follana, A.~Vaquero, G.~Di Carlo,
  JHEP {\bf 0908} (2009) 008.
%
  T.~Korzec, U.~Wolff,
  PoS LATTICE {\bf 2010} (2010) 029.
%
  P.N.~Meisinger, M.C.~Ogilvie,
  arXiv:1306.1495 [hep-lat].
  
\bibitem{prl}
  Y.~D.~Mercado, C.~Gattringer and A.~Schmidt,
  Phys.\ Rev.\ Lett.\  {\bf 111} (2013) 141601
  [arXiv:1307.6120 [hep-lat]].
  
\bibitem{z3}
  C.~Gattringer and A.~Schmidt,
  Phys.\ Rev.\ D {\bf 86} (2012) 094506
  [arXiv:1208.6472 [hep-lat]].
  
\bibitem{swa}
  Y.~D.~Mercado, C.~Gattringer and A.~Schmidt,
  Comput.\ Phys.\ Commun.\  {\bf 184} (2013) 1535
  [arXiv:1211.3436 [hep-lat]].

\bibitem{worm}
  N.~Prokof'ev and B.~Svistunov,
  Phys.\ Rev.\ Lett.\  {\bf 87} (2001) 160601.

\bibitem{Lang}
 K.~Jansen, J.~Jersak, C.B.~Lang, T.~Neuhaus, G.~Vones,
  Nucl.\ Phys.\ B {\bf 265} (1986) 129;
  Phys.\ Lett.\ B {\bf 155} (1985) 268.
  K.~Sawamura, T.~Hiramatsu, K.~Ozaki, I.~Ichinose,
  arXiv:0711.0818 [cond-mat.str-el].
  
\bibitem{cohen}
T.D.~Cohen,
  Phys.\ Rev.\ Lett.\  {\bf 91} (2003) 222001.

\end{thebibliography}
\end{document}